# Light amplification by stimulated emission from an optically pumped molecular junction in a scanning tunneling microscope

*K. Braun, A. M. Kern, X. Wang, H. Adler, H. Peisert, T. Chassé, D. Zhang, and A. J. Meixner*

*IPTC Universität Tübingen, Auf der Morgenstelle 18, 72076 Tübingen, Germany*

**Electromagnetic coupling between resonant plasmonic oscillations of two closely spaced noble metal particles can lead to a strongly enhanced optical near field in the gap between. This effect is widely used in surface- and tip-enhanced Raman scattering (TERS)[1-4] or for the enhancement of molecular luminescence.[5] Resonant surface plasmon amplification by stimulated emission from a nanometer-sized optical gain medium, similar to a laser, has recently been predicted and experimentally realized.[6,7] However, new discoveries in quantum plasmonics show that an upper limit is imposed to the field enhancement by the intrinsic nonlocality of the dielectric response of the metal and the tunneling of the plasmon's electrons through the gap.[8,9] Here, we introduce and experimentally demonstrate optical amplification and stimulated emission from a single optically pumped molecular tunneling junction of a scanning tunneling microscope (STM). The gap between a sharp gold tip and a flat gold substrate covered with a self-assembled monolayer (SAM) of 5-chloro-2-mercaptobenzothiazole (Cl-MBT) molecules forms an extremely small optical gain medium. When electrons tunnel from the Cl-MBT's highest occupied molecular orbital (HOMO) to the tip, holes are left behind. These can be repopulated by hot electrons induced by the laser-driven plasmon oscillation on the metal surfaces enclosing the cavity. Solving the laser-rate equations for this system shows that the repopulation process can be efficiently stimulated by the gap mode's near field, TERS scattering from neighboring molecules acting as an optical seed. Our results demonstrate how optical enhancement inside the plasmonic cavity can be further increased by a stronger localization via tunneling through molecules. We anticipate that stimulated emission from an STM junction will advance our fundamental understanding of quantum plasmonics and lead to new analytical applications. Furthermore, this concept represents the basis for novel ultra-small, fast, optically and electronically switchable devices and could find applications in high-speed signal processing and optical telecommunications.**

The emission of photons from the gap of a scanning tunneling microscope (STM) has been a focus of interest for more than twenty years[10] and has been used for acquiring spectroscopic information with ultra-high spatial resolution.[11] For pure metal surfaces[12,13] or organic monolayers adsorbed directly on a metal surface,[14] the emission of light originates predominantly from the radiative decay of localized surface plasmons (LSP) excited by inelastic electron tunneling (IET) as the direct luminescence of the molecules is quenched. If the molecules are decoupled from the metal surface by an ultra-thin dielectric layer, intrinsic molecular luminescence can be observed down to the single-molecule level, showing vibronic bands.[15,16] The efficiency of this photoemission is very low, typically one emitted photon per $10^5$-$10^6$ tunneling electrons.[10] In recent years, a different approach for ultra-high resolution optical spectroscopy has emerged. So-called tip enhanced Raman scattering (TERS),[3,17-19] or gap mode near-field optical microscopy, has attracted great interest as a means for local Raman or luminescence spectroscopy with nanometer spatial resolution. In this approach, the metal tip serves as an optical antenna and generates an electromagnetic near field in the gap, locally enhancing excitation. At the same time, the emission and radiation of photons into the far field[20] is enhanced due to the increased local density of optical states in the gap. Since efficient Raman scattering from molecules in the gap requires gap widths as short as one nanometer, electron tunneling is meanwhile routinely used to control the tip/sample distance.[4,21,22]

Here we show amplification of TERS and luminescence emission from an STM junction (Fig. 1a) by applying a bias voltage exceeding a threshold of $U_b > 1000$ mV. In the low bias-voltage range, i.e. for $|U_b| < 1000$ mV, the spectra (Fig. 1b and 1c) are almost independent of $U_b$ and one can observe the typical tip-enhanced Raman bands of Cl-MBT residing on a broad luminescence background. The prominent bands are the aromatic C-C stretch vibrations around 1600 cm$^{-1}$ and the aromatic C-H stretch vibrations at 2900 cm$^{-1}$. When $U_b$ exceeds the threshold of 1000 mV we observe an intensity increase of one order of magnitude affecting primarily the aromatic Raman bands around 1600 cm$^{-1}$, whereas for a negative bias voltage the intensity increases only by a factor of two and affects the whole spectrum (Fig. 1c). This behavior is fully reversible up to $U_b$=2000 mV (data see supplementary). The dependence on the incident laser power also exhibits a distinct nonlinear behavior with a threshold separating a low-gain regime and a high-gain regime (Fig.1d), both for continuous-wave (CW) and pulsed excitation (pulse length 100 ps, 80 MHz repetition rate). Both curves show the nonlinear onset at comparable equivalent powers and have the same slope in the high-gain regime (see supplementary). This indicates that the system has reached stationary conditions within 100 ps, as expected for typical plasmon relaxation times on the order of 100 fs.[23] When the laser irradiation is switched off light emission is three orders of magnitude weaker and occurs exclusively from inelastic tunneling at bias voltages $|U_b| \geq 1500$ mV (Fig. 2b).

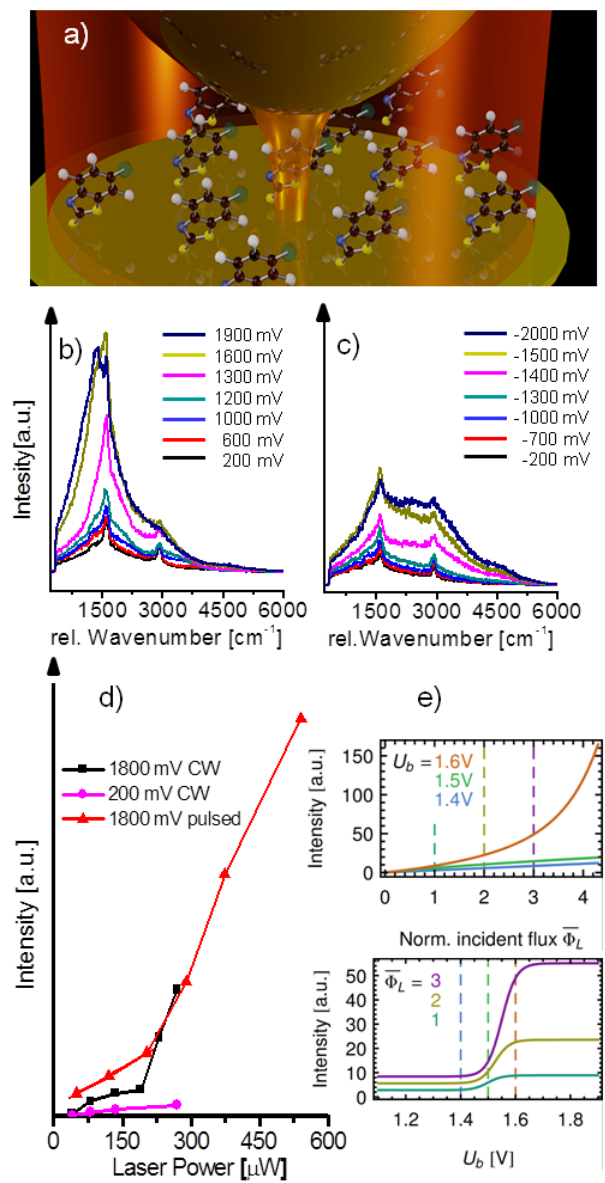

Fig. 1a) Tunneling junction consisting of a sharp gold that tip statically positioned above a gold substrate covered with a monolayer of chemisorbed Cl-MBT-molecules. Sequences optical spectra recorded from the tunneling as a function of the positive bias voltage (Fig. 1b) or negative bias voltage (Fig. 1c). (Fig. 1d) Spectrally integrated Stokes-shifted emission intensity as function of the incident laser power for continuous-wave (CW) and pulsed illumination (CW-equivalent power) showing for a distinct threshold between a low- and high gain regime. (Fig. 1e) Theoretical curves showing the emission rate $\gamma_{em}$ as a function of the incident laser photon flux $\Phi_L$ (upper case) and the bias voltage $U_b$ for descriptive values of the experimental parameters.

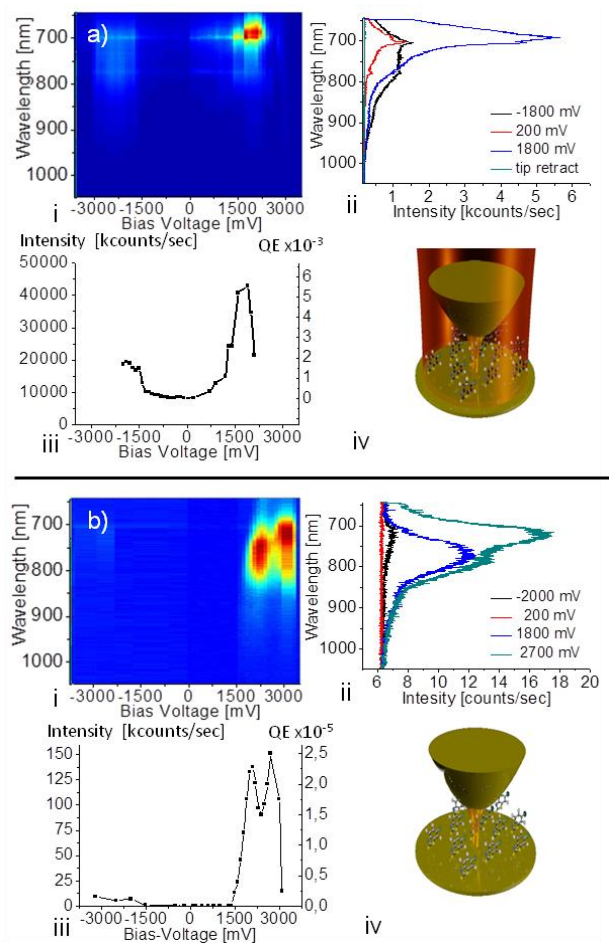

Fig. 2a) Tip-enhanced emission spectra from the tunneling junction excited at $\lambda = 632.8$nm with 250µW as a function of the bias voltage(i) with the respective spectrally integrated intensity trajectory (iii) as a function of bias voltage. (Fig 2b) Luminescence spectra excited by inelastic tunneling without laser illumination as a function of the bias-voltage (i). All spectra were recorded with the same tunneling current (1nA) and are normalized to 1s integration time. The corresponding quantum efficiencies (QEs) are defined as the number of Stokes-shifted photons per tunneling electron and can directly be compared for the two situations.

The spectra show the typical broad tip plasmon modes, which depend on the morphological details of the tip-sample junction[10] and look very similar to the emission bands observed from inelastic tunneling and radiative plasmon relaxation of a pure Au-tip/Au-sample junction (see Supplementary Fig. 3d). For negative bias voltages, on the other hand, the Cl-MBT junction exhibits only very weak photon emission in contrast to the Au/Au case, from which similar photon emission intensities can be observed for both polarities when $|U_b| \geq 1500$ mV (see Supplementary Fig. 3d). By comparing the intensity trajectories (Figs. 2a iii and Fig. 2b iii) one can see that the spectrally integrated intensity of the optical signal caused by inelastic tunneling alone is more than two orders of magnitude smaller than the bias voltage dependent increase of the optical signal of the laser illuminated junction, meaning that the increase of the optical signal in Figs. 1b) and 2a) is not only an additive effect but a true amplification when the bias voltage exceeds 1300 mV. It thus seems likely that the energy levels of the surface-bound molecules and the density of electronic surface states of the Au substrate along with their relative alignment play a paramount role in the observed optical emission and amplification processes. Indeed, ultraviolet photoelectron spectroscopy (UPS) measurements of a monolayer of Cl-MBT on Au reveal the energetic distribution of the HOMO of the surface-bound molecules with respect to the Fermi level of the Au surface at around -1.5 – 2.0 eV and the d-bands of Au below -2.0 eV (see Supplementary Fig. 2). These energy ranges coincide with the bias voltages for which we observe the step-like increases of the TERS signal and the laser-induced Au luminescence signal, vis. Figs. 1b), 2a) and 2b), respectively. To further elucidate the role of the tip's bias, we first consider the light emission observed without laser irradiation (Fig. 2b and supplementary Fig. 5d). For a pure Au/Au junction the emitted light originates from inelastic electron tunneling and the radiative decay of the consequently excited surface plasmons independent of polarity.[10-12,13,24] In contrast the Au/Cl-MBT/Au junction light emission induced by inelastic tunneling is one order of magnitude more efficient when the tip is positively charged than for a negatively charged tip. Obviously the surface bound molecules have influence on light emission by inelastic tunneling. For positive bias voltages $U_b \geq 1500$ mV, the sharp intensity increase seen in Fig. 2b) indicates that electrons can additionally (elastically) tunnel from the HOMO of the surface-bound molecules (at -1.5 – 2.0 eV) to the positively charged tip. For electrons near the Fermi energy recombining with a bound molecule's HOMO, the excess energy corresponds to the PL band at around 775 nm observed for $U_b \geq 1500$ mV (blue spectrum in Fig. 2b). When the bias voltage is further increased, $U_b \geq 2000$ mV, electrons from surface orbitals of the d-band begin to tunnel elastically to the tip, also giving rise to holes. The recombination of these holes with electrons from the Fermi level is then the source of the energetically higher, blue-shifted band at 725 nm (green spectrum in Fig. 2b). If the bias voltage is inverted (negative tip and positive sample) we observe only a very weak PL signal from the Cl-MBT-adsorbed sample. Evidently, electron-hole formation and radiative recombination is hindered by the inverted polarity. In contrast, for a bare Au/Au junction the tunneling-induced luminescence is only weakly polarity-dependent, with differences between both polarities due to the geometrical difference between the tip apex and the substrate surface. We thus draw the conclusion that electron-hole formation and radiative recombination occurs on very localized scale, only in those few surface-bound molecules that form a tunneling junction between the substrate and the tip. The source volume of the molecular PL is thus much smaller even than the plasmonic gap mode.

Illuminating the tip/sample junction in addition with a focused laser beam polarized along the tip's axis, a coupled surface plasmon oscillation in the tip and the underlying sample surface is induced, manifesting itself as a highly localized surface charge oscillation at the very apex of the tip and the Au surface below. Recently, it has been shown that plasmon-excited nanoparticles can be an efficient source of hot electrons[25,26]. Each surface plasmon quantum can either decay into a photon (via scattering) or into an electron-hole pair (EHP, via absorption) that can recombine under emission of a luminescence photon. This luminescence emission appears spatially collocated with the plasmons themselves,[27] implying that the hot electrons remain localized until their decay. New studies of very small gaps (< 1 nm) and nearly touching pure metal nanoparticles show that an upper limit is imposed to the field enhancement in the gap by the intrinsic nonlocality of the metal's dielectric response, which leads to electron spill-out and the tunneling of the plasmon-excited electrons through the gap.[8,9,28] In our experiments, the confinement limit of electron tunneling and electron-hole recombination is defined only by the individual surface-bound molecules.

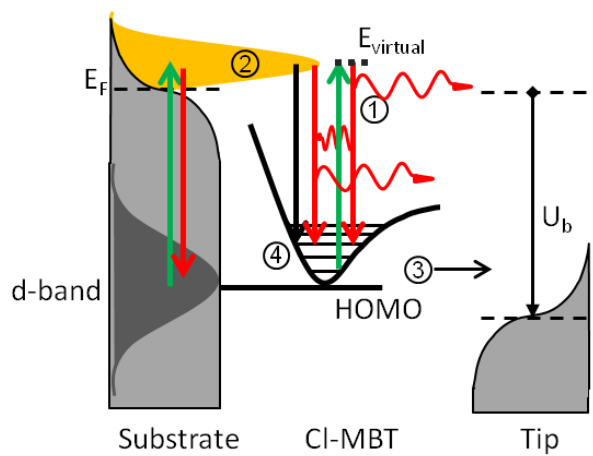

Fig. 3) Schematic energy level diagram of the gap/molecule hybrid system in a laser-illuminated tunneling junctio. Green arrows denote processes drawing energy from the incident laser field and red arrows represent emission or scattering processes. The incident radiation leads to (1) Raman scattering form the surface-bound molecules and (2) generates hot surface electrons with a non-thermal distribution indicated in yellow. Nonradiative processes are indicated by black arrows, i.e. elastic tunneling of electrons from the molecules' HOMO levels to the tip (3) when the Fermi-level of the tip falls below the HOMO-level of the closest surface-bound molecules and nonradiative recombination of hot electrons (4) with holes in the molecules repopulating the HOMO levels.

Having understood the individual processes involved, one can summarize the behavior of the Au/Cl-MBT/Au junction in an energy level scheme as represented in Fig. 3. Green arrows represent processes drawing energy from the incident pump-laser field, i.e. Raman scattering from the surface-bound molecules (1) and generation of hot electrons from the d-band (2). Plasmon excitation is not shown in this figure. When the Fermi level of the tip drops below the energy of a surface-bound molecule's HOMO energy, efficient elastic electron tunneling from the molecule closest to the tip sets in (3), leaving behind a hole in the molecule. This hole can then easily be refilled by a hot surface electron from the Au substrate (4), emitting a photon which can be observed by the detector. The rate of this optical process depends on two factors: the local density $\rho_e^*$ of hot surface electrons which is a function of the incident laser power, and the tunneling rate $k_t$ from the HOMO to the tip which is a function of the applied bias voltage.

The nonlinear increase of the emission intensity with increasing incident laser power is typical for a system in which a positive feedback process is taking place, directing a portion of the amplified Raman and PL emission back towards the substrate/molecule hybrid system. Considering the fact that the Raman scattering and PL emission processes originate in the very center of the tip/substrate gap, any generated photon will first couple to the gap mode before being scattered to the far field. While the gap mode's plasmon resonance is very broad, exhibiting a quality factor of only Q≈15, the resonantly stored energy is extremely well localized spatially, in a volume having an upper limit of approximately 4×4×1 nm$^3$ (see Supplementary Fig. 1). The feedback of the emitted energy back to the quantum system can thus be very efficient, still. A stimulation of the radiative decay of hot electrons into the molecule's emptied HOMO level can explain the nonlinear dependence of the emission intensity on the incident laser power.

The optical system comprising the gap mode, plasmonic feed-back and the substrate/molecule hybrid system is reminiscent of a novel nanoscale laser, containing an extremely small and fast optical gain medium exhibiting a population inversion, i.e. the hot-electron recombination of the molecule's depleted HOMO level, bi-directionally coupled to a resonator, i.e. the plasmonic gap mode. In this case, population inversion occurs between an empty HOMO-level and the higher lying energy level of hot electrons and thus gain is expected under the very conditions for which enhanced emission can be observed experimentally: depletion of the HOMO via elastic tunneling to the tip and generation of hot electrons by irradiating the plasmonic system with incident laser light. TERS scattering from neighboring molecules acts as an optical seed signal in the cavity mode and is amplified in the molecular tunneling junction by stimulated emission from hot electrons recombining with holes in the surface-bound molecules. This leads to the spectral narrowing of the amplified signal (Fig. 1b) in the metal-molecule-metal junction for a positive bias voltage above threshold. For a negative bias voltage electrons can tunnel directly from the tip to the metal substrate without the creation of holes in the HOMO-level of the surface bound molecules. In this case the molecules rather form a thin insolating layer hindering light emission from inelastic tunneling. The behavior of the molecule/gap hybrid system can be described with coupled rate-equation model for the population $P_0$ of the molecule's HOMO level and the gap mode's energy density $\rho_\phi$ (see Supplementary) reproducing the nonlinear increase of the emitted signal as a function of the incident laser power and the bias voltage (Fig. 1e) for the experimental parameters underlying our experiment.

Our work demonstrates experimentally and explains theoretically a novel concept for an ultrasmall plasmonic laser based on a single hybrid plasmon/molecule/plasmon tunneling junction and opens new prospects for novel ultra-small (for example based on single molecules), fast, optically and electronically switchable devices that could find applications in high-speed signal processing and optical telecommunications.

**Methods**

All measurements were performed with a fully homebuilt parabolic mirror scanning near-field optical microscope[29,30] equipped with an also homebuilt STM scanner. A 200 nm thick gold layer was evaporated on a silicon wafer. These substrates were then dipped for 10 minutes in a $10^{-2}$ M solution of Cl-MBT (Sigma Aldrich, 90% technical grade) molecules in Uvasol methanol. Afterwards, the samples were carefully rinsed with methanol and dried in a clean-air box. The gold tips were prepared by electrochemical etching in a fuming HCl solution(Sigma Aldrich Art.-Nr. 84415). A helium-neon laser (Melles Griot 25LHP991-230) at 632.8 nm (250 µW) was used as a CW excitation source. Integration time per individual spectrum was between 1 and 3 s. To compare the spectra with different integration times we normalized all spectra to one second and calculate the QE according to the number of electrons tunneling in this time. As a second excitation source we used a pulsed diode laser (Picoquant LDH-D-C-640) at 635 nm (up to 270 µW). During a spectra series the tip was positioned statically above the surface and the tunneling current was kept constant at 1 nA.

**References**

1	Liu, Z. *et al.* Revealing the molecular structure of single-molecule junctions in different conductance states by fishing-mode tip-enhanced Raman spectroscopy. *Nat Commun* **2**, 305, (2011).

2	Michaels, A. M., Jiang & Brus, L. Ag Nanocrystal Junctions as the Site for Surface-Enhanced Raman Scattering of Single Rhodamine 6G Molecules. *J. Phys. Chem. B* **104**, 11965-11971, (2000).

3	Stockle, R. M., Suh, Y. D., Deckert, V. & Zenobi, R. Nanoscale chemical analysis by tip-enhanced Raman spectroscopy. *Chem. Phys. Lett.* **318**, 131-136, (2000).

4	Zhang, R. *et al.* Chemical mapping of a single molecule by plasmon-enhanced Raman scattering. *Nature* **498**, 82-86, (2013).

5	Kinkhabwala, A. *et al.* Large single-molecule fluorescence enhancements produced by a bowtie nanoantenna. *Nat Photon* **3**, 654-657, (2009).

6	Bergman, D. J. & Stockman, M. I. Surface Plasmon Amplification by Stimulated Emission of Radiation: Quantum Generation of Coherent Surface Plasmons in Nanosystems. *Phys. Rev. Lett.* **90**, 027402, (2003).

7	Noginov, M. A. *et al.* Demonstration of a spaser-based nanolaser. *Nature* **460**, 1110-1112, (2009).

8	Ciracì, C. *et al.* Probing the Ultimate Limits of Plasmonic Enhancement. *Science* **337**, 1072-1074, (2012).

9	Savage, K. J. *et al.* Revealing the quantum regime in tunnelling plasmonics. *Nature* **491**, 574-577, (2012).

10	Rossel, F., Pivetta, M. & Schneider, W.-D. Luminescence experiments on supported molecules with the scanning tunneling microscope. *Surf. Sci. Rep.* **65**, 129-144, (2010).

11	Berndt, R., Gimzewski, J. K. & Schlittler, R. R. Enhanced photon emission from the STM: a general property of metal surfaces. *Ultramicroscopy* **42–44, Part 1**, 355-359, (1992).

12	Berndt, R., Gimzewski, J. K. & Johansson, P. Inelastic tunneling excitation of tip-induced plasmon modes on noble-metal surfaces. *Phys. Rev. Lett.* **67**, 3796-3799, (1991).

13	Johansson, P., Monreal, R. & Apell, P. Theory for light emission from a scanning tunneling microscope. *Phys. Rev. B* **42**, 9210-9213, (1990).

14	Berndt, R. *et al.* Photon Emission at Molecular Resolution Induced by a Scanning Tunneling Microscope. *Science* **262**, 1425-1427, (1993).

15	Qiu, X. H., Nazin, G. V. & Ho, W. Vibrationally Resolved Fluorescence Excited with Submolecular Precision. *Science* **299**, 542-546, (2003).

16      Wu, S. W., Nazin, G. V. & Ho, W. Intramolecular photon emission from a single molecule in a scanning tunneling microscope. *Phys. Rev. B* **77**, 205430, (2008).

17      Hartschuh, A., Pedrosa, H. N., Novotny, L. & Krauss, T. D. Simultaneous Fluorescence and Raman Scattering from Single Carbon Nanotubes. *Science* **301**, 1354-1356, (2003).

18      Hayazawa, N., Inouye, Y., Sekkat, Z. & Kawata, S. Metallized tip amplification of near-field Raman scattering. *Opt. Commun.* **183**, 333-336, (2000).

19      Raschke, M. B. & Lienau, C. Apertureless near-field optical microscopy: Tip-sample coupling in elastic light scattering. *Appl. Phys. Lett.* **83**, 5089-5091, (2003).

20      Paolo, B., Jer-Shing, H. & Bert, H. Nanoantennas for visible and infrared radiation. *Rep. Prog. Phys.* **75**, 024402, (2012).

21      Stadler, J., Schmid, T. & Zenobi, R. Nanoscale Chemical Imaging Using Top-Illumination Tip-Enhanced Raman Spectroscopy. *Nano Lett.* **10**, 4514-4520, (2010).

22      Steidtner, J. & Pettinger, B. Tip-enhanced Raman spectroscopy and microscopy on single dye molecules with 15 nm resolution. *Phys. Rev. Lett.* **100**, 236101, (2008).

23      Mark, I. S. The spaser as a nanoscale quantum generator and ultrafast amplifier. *J. Opt.* **12**, 024004, (2010).

24      Gimzewski, J. K., Sass, J. K., Schlitter, R. R. & Schott, J. Enhanced Photon Emission in Scanning Tunnelling Microscopy. *Europhys. Lett.* **8**, 435, (1989).

25      Knight, M. W., Sobhani, H., Nordlander, P. & Halas, N. J. Photodetection with Active Optical Antennas. *Science* **332**, 702-704, (2011).

26      Mukherjee, S. *et al.* Hot Electrons Do the Impossible: Plasmon-Induced Dissociation of H2 on Au. *Nano Lett.* **13**, 240-247, (2012).

27      Fleischer, M. *et al.* Three-dimensional optical antennas: Nanocones in an apertureless scanning near-field microscope. *Appl. Phys. Lett.* **93**, 111114, (2008).

28      Esteban, R., Borisov, A. G., Nordlander, P. & Aizpurua, J. Bridging quantum and classical plasmonics with a quantum-corrected model. *Nat Commun* **3**, 825, (2012).

29      Sackrow, M., Stanciu, C., Lieb, M. A. & Meixner, A. J. Imaging nanometre-sized hot spots on smooth Au films with high-resolution tip-enhanced luminescence and Raman near-field optical microscopy. *Chemphyschem* **9**, 316-320, (2008).

30      Zhang, D. *et al.* Nanoscale Spectroscopic Imaging of Organic Semiconductor Films by Plasmon-Polariton Coupling. *Phys. Rev. Lett.* **104**, 056601, (2010).

**Acknowledgement**

This work was supported by the Deutsche Forschungsgemeinschaft, Grant 1600/5-3 and SPP1391. We acknowledge the Helmholtz-Zentrum Berlin - electron storage ring BESSY II for provision of synchrotron radiation and for financial travel support.